%% file: main.tex
\documentclass[conference]{IEEEtran}
\IEEEoverridecommandlockouts
\usepackage{cite}
\usepackage{amsmath,amssymb,amsfonts}
\usepackage{algorithmic}
\usepackage{graphicx}
\usepackage{textcomp}
\usepackage{xcolor}
\usepackage{comment}
\usepackage{caption} 
\usepackage{subcaption} 
\usepackage{array}
\usepackage{url}
\usepackage{tabularx}
\newcolumntype{C}{>{\centering\arraybackslash}X}
\newcolumntype{M}{>{\centering\arraybackslash}m{1.78cm}}

\def\BibTeX{{\rm B\kern-.05em{\sc i\kern-.025em b}\kern-.08em
T\kern-.1667em\lower.7ex\hbox{E}\kern-.125emX}}

\allowdisplaybreaks

\begin{document}


\title{Integrating Health Sensing into Cellular Networks: Human Sleep Monitoring Using 5G Signals}

\author{
    \IEEEauthorblockN{Ruxin Lin, Peihao Yan, Jie Lu, Qijun Wang, and Huacheng Zeng}
    \IEEEauthorblockA{{Department of Computer Science and Engineering, Michigan State University, USA}}\vspace{-0.35in} 
    \thanks{This work was supported in part by NSF Grants CNS-2312448 and ECCS-2434001.}
}


\maketitle

\begin{abstract}
Cellular networks offer a unique opportunity to enable device-free and wide-area health monitoring by exploiting the sensitivity of radio-frequency (RF) propagation to human physiological activities. In this paper, we present the first experimental study of human sleep monitoring using realistic 5G signals collected from commercial cellular infrastructure. We investigate a practical scenario in which a smartphone is placed near a bed, and a 5G base station periodically configures uplink sounding reference signal (SRS) transmissions to obtain fine-grained channel state information (CSI). Leveraging uplink CSI measurements, we design a lightweight signal processing pipeline for respiration rate estimation and a CNN model for sleep body movement classification. Through extensive experiments conducted on an indoor private 5G network, our system achieves over 91.2\% accuracy in respiration rate estimation and 85.5\% accuracy in sleep movement classification. 
\end{abstract}

\begin{IEEEkeywords}
5G/6G, health monitoring, integrated sensing and communication (ISAC), respiration detection
\end{IEEEkeywords}

\section{Introduction} 
\input{introduction}

\section{Related Work}
\input{related_work}
\section{Understanding 5G CSI for Sensing}
\input{preliminary}

\section{Design}

\input{design}

\section{Experimental Evaluation}
\input{evaluation}

\section{Conclusion}

In this paper, we studied human respiration and body movement monitoring during sleep using 5G signals. We proposed a signal processing pipeline for respiration rate estimation and a CNN-based classifier for sleep movement recognition. More importantly, we demonstrated through experiments on an indoor 5G network that the proposed approach, albeit simple, can achieve over 91.2\% accuracy in respiration rate detection and 85.5\% accuracy in movements classification. We hope this study will spur further research on healthcare applications leveraging existing and future 5G/6G infrastructure.

\bibliographystyle{IEEEtran}
\bibliography{references}
\end{document}

%% file: introduction.tex
Cellular signals can be leveraged for health detection by exploiting the sensitivity of radio-frequency (RF) propagation to human physiological activities. Subtle chest movements caused by respiration, as well as larger body motions, introduce measurable variations in the amplitude, phase, and Doppler characteristics of cellular signals reflected from or transmitted through the human body. These effects can be captured through fine-grained physical-layer measurements such as channel state information (CSI), which are already available in modern cellular systems. By applying advanced signal processing techniques and AI-driven inference models, respiration rates and movement patterns can be extracted without requiring any wearable devices. 

Furthermore, cellular-based health sensing addresses critical emerging needs that cannot be effectively met by other sensing technologies. Consider a scenario where an individual attempts to check on the health condition of his/her elderly parent who lives alone but cannot be reached by phone or any other means of communication. In such cases,  wearable devices or in-home WiFi sensing may not work due to their service locality or lack of deployment. Cellular-based sensing, however, can leverage the existing network connection associated with the parent's cellphone to passively infer respiration and movement status, offering a potentially life-saving capability. 
In another scenario, law enforcement agencies may want to assess the health condition of a subject when only his/her phone number is available. Unlike WiFi-based sensing, which requires proximity and prior infrastructure access, cellular-based sensing enables long-distance and wide-area detection through a large-scale network infrastructure. 
For both scenarios, cellular-based sensing may represent the only viable solution, as it operates independently of local network access, user participation, or additional hardware. 

Despite its significant potential, cellular-based health detection remains largely underexplored. A primary barrier is the limited accessibility of CSI data in cellular networks, as mobile network operators are generally reluctant to share physical-layer data required to evaluate the flexibility and performance of emerging sensing solutions. Recently, the advent of Open Radio Access Network (O-RAN) architectures, together with the growing momentum of integrated sensing and communication (ISAC), has created new opportunities to investigate cellular sensing through experimental studies. As cellular infrastructure evolves from tightly coupled, proprietary 5G systems toward open and AI-native 6G architectures based on O-RAN principles, the exposure of standardized open interfaces (e.g., E2) not only enables data-driven, learning-based network optimization but also facilitates the integration of native sensing services, positioning cellular networks as a unified communication-and-sensing platform.

\begin{figure}
    \centering
    \includegraphics[width=\linewidth]{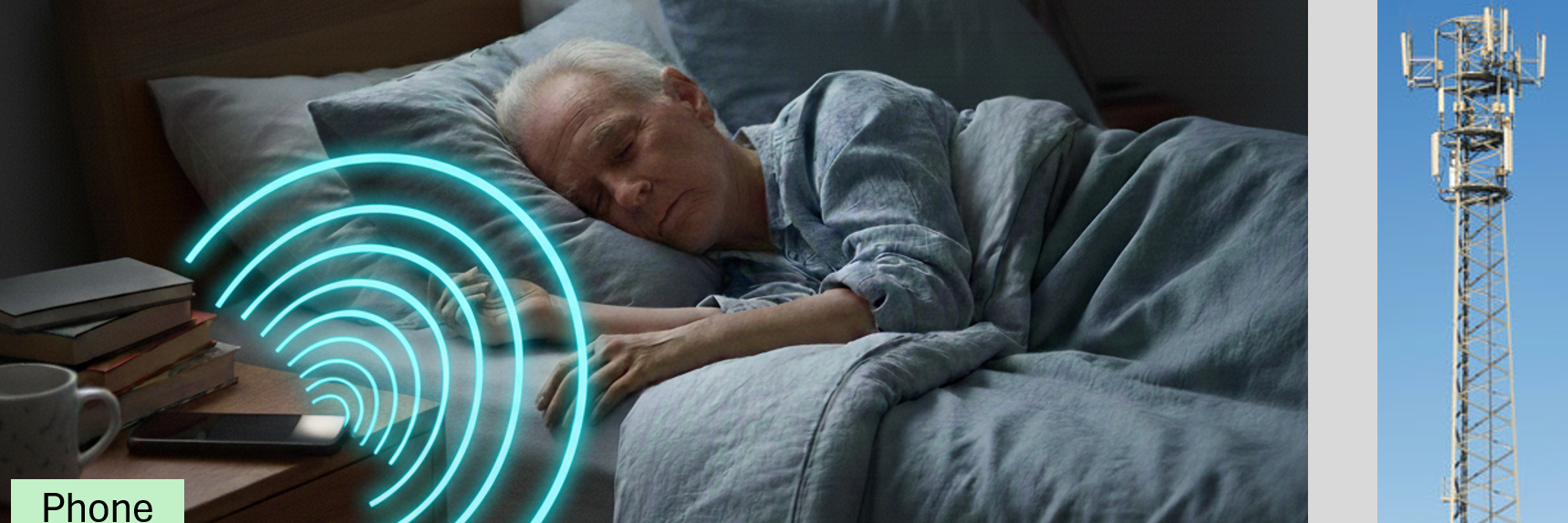}
    \caption{Human sleep monitoring using 5G networks.}\vspace{-0.2in}
    \label{fig:teaser2}
\end{figure}

In this paper, we investigate the feasibility of integrating human health sensing into NextG cellular networks. We consider the scenario illustrated in Fig.~\ref{fig:teaser2}, where an individual places a smartphone in close proximity to the bed during sleep. By leveraging its centralized control capability, the cellular base station configures the phone to periodically transmit uplink sounding reference signals (SRS) and estimates the corresponding uplink channel state information (CSI). Using the measured uplink CSI, the base station further infers physiological and behavioral information, including respiration rate estimation and body movement detection and classification during sleep. This information directly reflects the sleep quality of the individual.

Toward the 5G CSI-based health sensing, we face two challenges  that need to be addressed. 
The first challenge involves the interference from other moving objects in the environment. 
Cellular human sensing is often subject to interference from surrounding objects and dynamic environments, which can obscure subtle respiration signatures of the target subject. 
To address this question, we found via experiments that the measured CSI is most sensitive to the moving objects in the proximity of transmitter (phone) and receiver (base station), but not sensitive to those moving objects in the middle. 
Additionally, there is typically not much movement nearby the phone during sleep.
Based on these observations, we design a lightweight signal processing pipeline that utilizes the ratio of CSI samples from different antennas for respiration rate estimation. 

The second challenge lies in the irregular and diverse nature of human body movements during sleep, which complicates reliable movement detection and classification. To address this issue, we propose a CNN-based model that takes sanitized CSI amplitude measurements as input for sleep movement classification. Despite its simple architecture, the proposed model turns out to be effective in distinguishing predefined body movement patterns.

We implement the proposed sleep monitoring system on an indoor 5G network composed of commercial 5G radio units and smartphones. CSI is collected directly at the 5G base station, and respiration estimation and body movement detection are evaluated across five different deployment locations. Experimental results demonstrate that the proposed approach, although lightweight, achieves over 91.2\% accuracy in respiration rate estimation and 85.5\% accuracy in body movement classification during sleep, validating its effectiveness for practical cellular-based health sensing. To the best of our knowledge, this is the first work that demonstrates sleep respiration and body movement detection using  5G signals.

%% file: related_work.tex
While wearable devices such as smartwatches have been widely adopted for physiological monitoring, they are often limited by user discomfort, compliance issues, and maintenance overhead. Recently, RF sensing has attracted increasing interest as a non-contact alternative for human health monitoring.
Generally, they can be classified to two categories.

\smallskip

\noindent
\textbf{WiFi CSI-based Human Sensing.}
CSI in WiFi networks has been extensively explored for sensing applications, including localization \cite{10.1145/2829988.2787487}, human activity recognition \cite{10.1145/3241539.3241570,song2024siwis}, gesture and posture prediction \cite{10.1145/2500423.2500436}, sleep movement monitoring \cite{8399492,10858664}, and vital sign detection \cite{10.1145/3351279}. While WiFi CSI has demonstrated impressive performance in detecting routine human activities and physiological motions, its application scenarios differ significantly from CSI-based sensing in cellular networks. In particular, 5G networks provide much larger coverage than WiFi, enabling long-distance human activity monitoring and supporting many applications that cannot be supported by WiFi.

\smallskip

\noindent
\textbf{Cellular CSI-based Health Sensing.}
Compared to WiFi CSI sensing, research on CSI-based sensing in 5G networks remains limited. One primary reason is that most existing 5G systems are proprietary, and mobile network operators (MNOs) are generally reluctant to share low-layer signal data for research purposes. Recently, the emergence of O-RAN architectures has significantly improved the openness of cellular networks, leading to growing interest in cellular sensing research.
In parallel, ISAC has emerged as a key technology in 5G and beyond, enabling various sensing applications such as localization, target detection, and environmental sensing by exploiting the high bandwidth, massive MIMO, and beamforming capabilities of cellular infrastructure.

Despite these advances, human health sensing using 5G signals remains largely unexplored. Several pioneering studies \cite{10888046,9985013} have investigated sleep-related sensing tasks, such as respiration monitoring, in so-called cellular settings. However, these works simply use generic OFDM signals for sensing but do not involve real commercial 5G equipment or operational 5G networks. This work fills this gap by experimentally examining the feasibility and performance of human sleep monitoring using 5G signals collected from commercial base stations.

%% file: preliminary.tex
\subsection{Uplink CSI Measurement in 5G Networks}

\begin{figure}[t]
    \centering
    \includegraphics[
    trim=0 0 0 0,
    clip,
    width=\linewidth,]{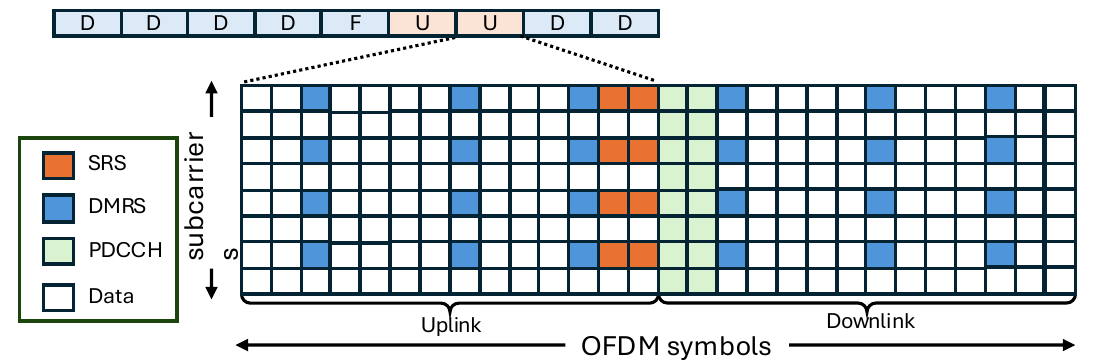}
    \caption{5G NR slot structure with SRS and DMRS}\vspace{-0.2in}
    \label{fig:srsframe}
\end{figure}

Uplink channel sounding is a fundamental function in 5G cellular networks, enabling reliable communication, efficient resource allocation, and emerging sensing applications. Unlike Wi-Fi networks,
5G networks adopt a centralized architecture based on a master-slave model. In this architecture, the base station serves as the central controller and orchestrates all uplink and downlink transmissions within the cell.

In a 5G system, UE strictly follows the scheduling and control commands issued by the base station. This centralized control allows the base station to configure each UE to transmit dedicated uplink reference signals, known as Sounding Reference Signals (SRS). By receiving these SRS transmissions, the base station can accurately estimate the uplink CSI, which captures the time-frequency-spatial characteristics of the wireless propagation environment. Importantly, the transmission of SRS can be explicitly scheduled and parameterized by the base station, including bandwidth, periodicity, and transmission power, enabling periodical channel measurements (e.g., one CSI per 20ms).
Therefore, SRS serves as a reliable and practical signal source for passive wireless sensing, forming the foundation of the proposed respiration and heart rate detection framework in this work.

Fig.~\ref{fig:srsframe} shows the SRS in an instance of 5G NR slot structure, where uplink and downlink transmissions are multiplexed across OFDM symbols. DMRS is embedded with data for channel estimation, SRS occupies dedicated uplink symbols for channel sounding, and PDCCH carries downlink control information.
For sensing purpose, we use SRS for uplink CSI measurement.

CSI characterizes how wireless signals propagate from the transmitter to the receiver, reflecting the combined effects of path loss, multipath propagation, reflection, scattering, and human interference in the environment. CSI has complex valued measurements over multiple subcarriers and antenna ports.
5G communications use OFDM modulation. 
Consider the uplink SRS transmission.
Denote $Y(k, t)$ as the received signal at the base station on subcarrier $k$ in frame $t$. 
Denote $\tilde{X}(k, t)$ as the original signal symbol transmitted by a UE. 
Then, the base station can estimate the UE's uplink channel by: 
\begin{equation}
    H(k,t) = \frac{Y(k, t)}{\tilde{X}c(k, t)},
\end{equation}
This CSI sample sequence, $H(k, t)$ with $t =1, 2, \dots$, captures the fine-grained temporal evolution of the wireless channel and provides a rich sensing modality beyond traditional RSS-based approaches.

\subsection{CSI for Sleep Monitoring}
One natural question regarding human sleep sensing is about the influence of other moving objects on the CSI measurement. 
In theory, CSI is affected not only by the physiological motions of a sleeping individual but also by any movement of objects along the signal propagation paths. Consequently, motion occurring anywhere within the 5G cell (e.g., people walking, doors opening, or other environmental changes) could affect the detection of  respiration and small body motions during sleep.

We address this challenge based on two observations. First, our experimental measurements show that only moving objects in close proximity to either the smartphone or the base station produce significant perturbations in the measured CSI. In contrast, moving objects located in the middle of the propagation path have a negligible impact on CSI, as their contributions are attenuated by distance and averaged out across multipath components. This observation is consistent with prior findings reported in the literature, which indicate that CSI sensitivity is dominated by motion near the transmitter or receiver.

\begin{figure}
    \centering
    \includegraphics[width=1\linewidth]{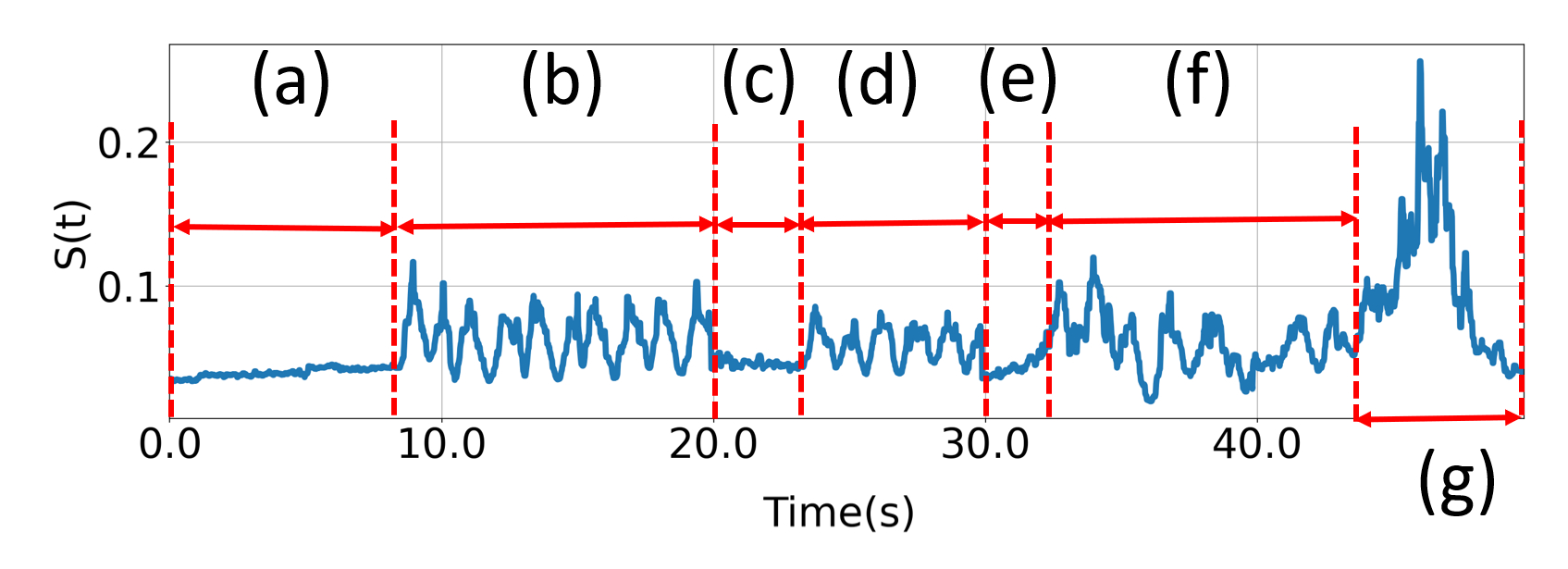}\vspace{-0.1in}
    \caption{Influence of moving objects on CSI when the moving objects are at different locations. $S(t)$ is CSI amplitude variance over time, i.e., $S(t) = \sum_{\forall k}|H(k, t) - H(k, t-1)|$.}
    \label{fig:interference}
\end{figure}

Fig.~\ref{fig:interference} shows the experimental observation of the measured CSI when the moving object (an interferer) is at different locations. 
In the figure, segment (a) shows the case without interference.
When no interference is present, subject motion induces clear CSI variations (segments (a)–(b)).
Outdoor interference introduces only minor fluctuations and does not obscure subject motion (segments (c)–(d)).Indoor interference away from the direct UE–RU path has limited impact (segments (e)–(f)). In contrast, when the interferer is very close to the UE (within 0.5 m), subject-induced CSI variations are largely masked (segment (g)).


Second, our sleep monitoring application targets nighttime periods.
During these periods, there are rarely moving objects in close proximity to the sleeping subject. Moreover, 5G base stations are mounted at elevated locations, making them unlikely to be affected by transient obstructions or nearby motion. Leveraging these observations, our system design focuses on isolating CSI variations induced by physiological motion, enabling the extraction of meaningful sleep-related features from measured 5G CSI despite the large spatial coverage of cellular networks.

%% file: design.tex
In this section, we present our approach for using 5G CSI measurements to estimate an individual’s respiration and body movements during sleep. Specifically, we aim to estimate the subject’s respiration rate and classify body movements into categories such as body turning, leg movement, arm movement, and sitting up.



\subsection{CSI Feature Extraction}

A 5G base station can enable uplink channel sounding by configuring a phone or other user devices to transmit SRS via over-the-air control signaling. The base station can further configure the time interval and bandwidth of the channel sounding process. At the base station, the measured CSI is denoted as
$H(i, k, t)$,
where $i$ is the antenna index (e.g., \(1 \le i \le 4\)), $k$ is the valid subcarrier index (e.g., \(1 \le k \le 800)\), and $t$ is the CSI sample index over time, with \(t = 1, 2, \dots\).

Both the amplitude and phase of the measured CSI carry information that reflects changes and movements in the surrounding environment. In theory, CSI phase is more sensitive to fine-grained human and object motions than CSI amplitude. For example, a displacement on the order of one-tenth of the carrier wavelength (e.g., approximately 0.86 cm at 3.5 GHz in the n78 band) can be captured through phase variations. However, such phase information can be reliably exploited only in monostatic sensing systems. In contrast, 5G communication systems operate in a bistatic configuration. The physical separation between the transmitter (UE) and the receiver (base station) introduces timing, frequency, and phase offsets in the measured CSI, resulting in phase misalignment that cannot be fully corrected. Consequently, the raw CSI phase cannot be directly used to infer human or object movement distances.

To leverage phase-related information in the presence of random phase offsets, we adopt normalized CSI as the input for human sleep detection. Specifically, we compute
\begin{equation}
    \tilde{H}(i, k, t) = \frac{H(i, k, t)}{H(0, k, t)},
\end{equation}
for \(t = 1, 2, \dots\).
The normalized CSI is then used to generate the two data sets for sleep detection, i.e., 
\begin{equation}
\left\{
\begin{array}{l}
A(i, k, t) = |\tilde{H}(i, k, t)|, \\
\phi(i, k, t) = \operatorname{unwrap}(\angle(\tilde{H}(i, k, t))).
\end{array}
\right.
\end{equation}

\begin{figure}[t]
    \centering

    \begin{subfigure}[t]{\linewidth}
        \centering
        \includegraphics[width=\linewidth]{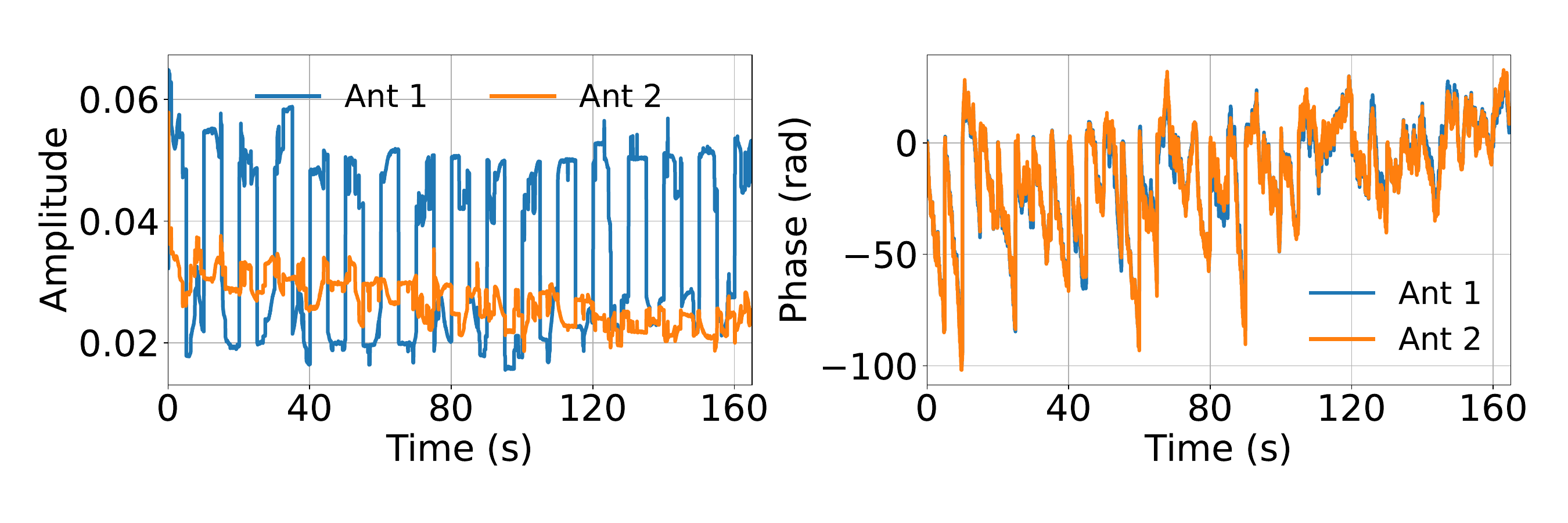}\vspace{-0.15in}
        \caption{Raw measured CSI.}
        \label{fig:raw_csi}
    \end{subfigure}

    \vspace{0.1em} 
    \begin{subfigure}[t]{\linewidth}
        \centering
        \includegraphics[width=\linewidth]{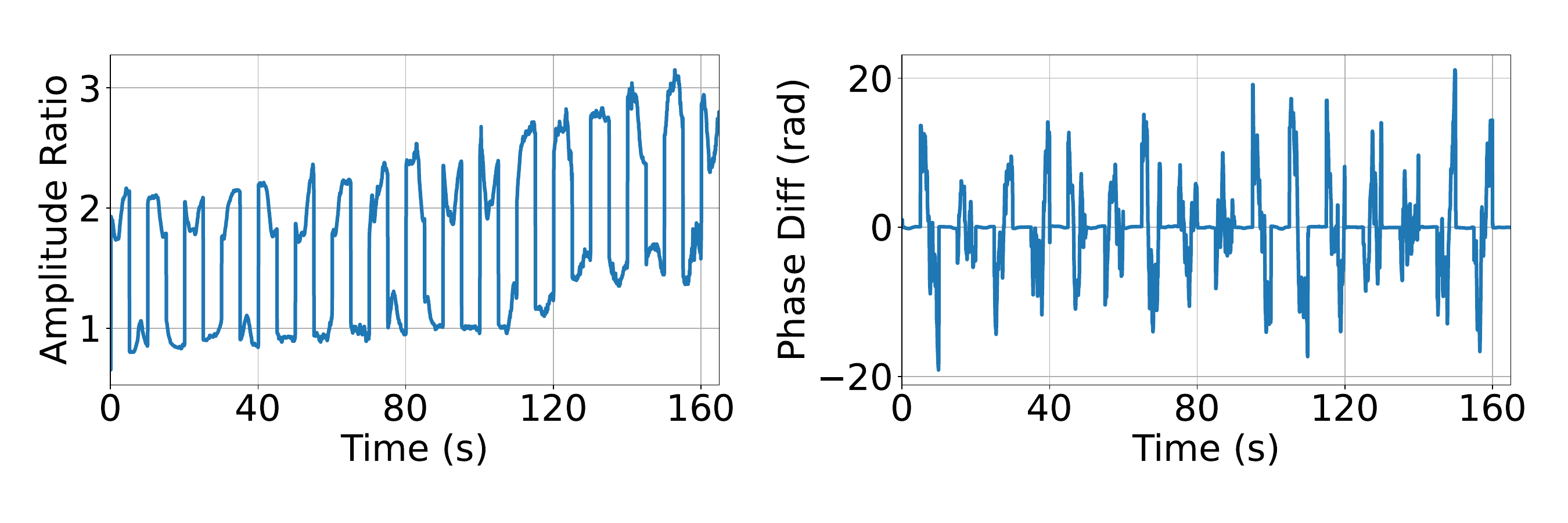}\vspace{-0.15in}
        \caption{Normalized CSI.}
        \label{fig:normalized_csi}
    \end{subfigure}

    \begin{subfigure}[t]{\linewidth}
        \centering
        \includegraphics[width=\linewidth]{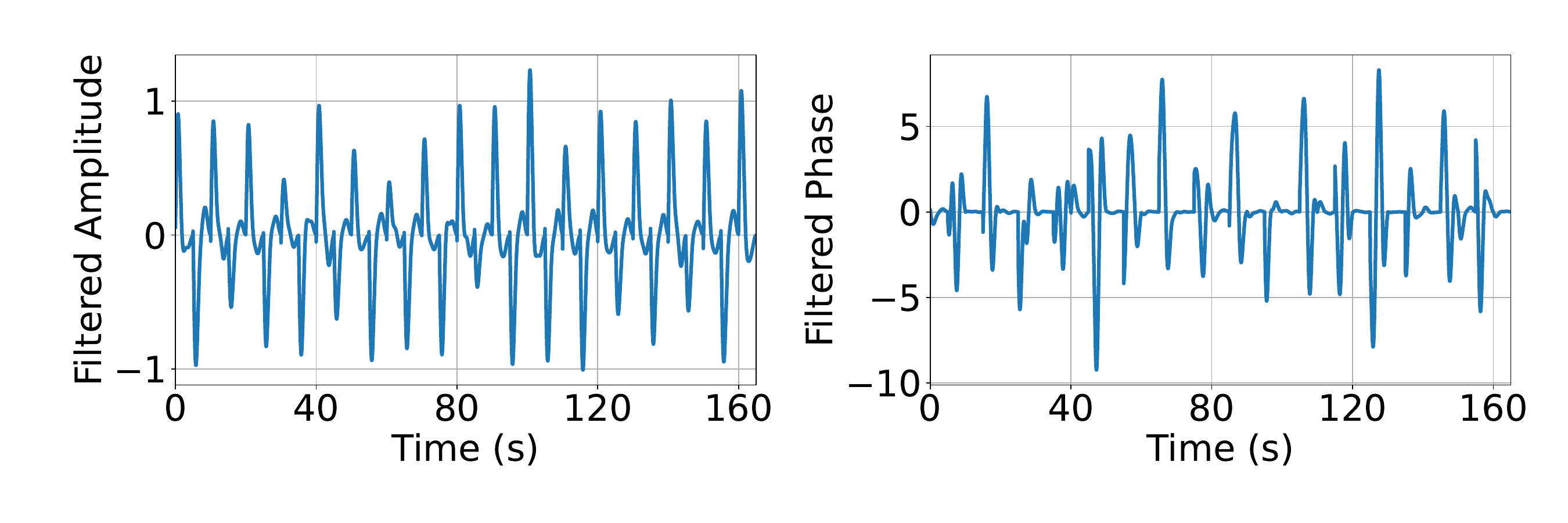}\vspace{-0.15in}
        \caption{Filtered CSI.}\vspace{-0.2in}
        \label{fig:processed_csi}
    \end{subfigure}
    
    \caption{An instance of measured raw CSI and normalized CSI.}\vspace{-0.2in}
    \label{fig:example}
\end{figure}

Fig.~\ref{fig:example}(a) presents a sample of the measured CSI in an indoor 5G network using a commercial radio unit and a Motorola smartphone. 
The phone was placed in a setting as shown in Fig.~\ref{fig:teaser2}, and there are no other object movements. It can be seen that the CSI phase appears random due to the uncontrolled phase offset between the smartphone and the base station. Fig.~\ref{fig:example}(b) shows the normalized CSI amplitude and phase, both of which exhibit much clearer patterns corresponding to respiration.

\subsection{Respiration Rate Estimation}

Based on our experiments, the normalized CSI exhibits a clear respiration-related pattern when no other moving objects are present in the vicinity of the smartphone. In contrast, when nearby people are moving, the CSI fluctuates significantly, and these variations can overwhelm the subtle changes induced by respiration. Fortunately, during nighttime sleep, human activity in the surrounding environment is typically minimal. Moreover, as demonstrated in prior experiments, moving objects at larger distances have a limited impact on the measured CSI.

To improve respiration estimation, we leverage the fact that human respiration typically lies within the 0.1--0.5 Hz frequency band, even when accounting for inter-subject variability and abnormal breathing patterns. Based on this observation, we apply a bandpass filter to the CSI amplitude and phase, respectively, with a bandpass of 0.1--0.5 Hz. This filtering suppresses low-frequency components caused by static objects as well as high-frequency components associated with noise and residual motion. Among all subcarriers, the one exhibiting the highest temporal stability is selected for respiration rate estimation.

Fig.~\ref{fig:processed_csi} illustrates the filtered CSI amplitude and phase obtained from the normalized CSI shown in Fig.~\ref{fig:normalized_csi}. The respiration frequency is estimated using peak detection in the time domain.
Although both CSI amplitude and phase can be used to detect chest motion from respiration, their robustness varies across environments. In practice, CSI amplitude is more resilient to hardware impairments and phase noise, while CSI phase can capture finer displacement variations but is more vulnerable to environmental interference and residual motion.
To combine respiration estimates from CSI amplitude and phase, we introduce a signal-quality–aware selection strategy. Let $x_a(t)$ and $x_p(t)$ denote the bandpass-filtered amplitude and phase signals, respectively.


To quantify the reliability of each modality, we define a spectral concentration metric:
\[
Q_{\text{spec}}^{(m)} =
\frac{|X_m(f_m^\ast)|}{\sum_{f \in \mathcal{B}} |X_m(f)|},
\quad m \in \{a,p\},
\]
where $X_m(f)$ is the Fourier transform of the filtered signal and $f_m^\ast$ denotes the dominant frequency.

The final respiration rate is selected from the modality with higher signal quality:
\[
\hat{f}_{\text{resp}} =
\arg\max_{m \in \{a,p\}} Q_{\text{spec}}^{(m)}.
\]
This lightweight strategy exploits the complementary robustness of amplitude and phase without introducing additional model parameters.






\subsection{Sleep Movement Classification}

In addition to respiration, the movement of sleep body and limb is another indicator related to the sleep quality. 
Unlike respiration, which is periodical activity, body movement is bursty and generates irregular CSI change patterns and has different time durations. 
As such, a model-based signal processing may not be competent. 
Therefore, we resort to a learning-based approach to detecting and classifying movements.



Different from respiration that involves subtle movement of human chest, body movement involves large-scale human body/limb displacement. 
Therefore, we use CSI amplitude only for this task. 
Specifically, we first process the CSI amplitude to reduce inter-sample scaling variation and mitigate residual automatic gain control (AGC) effects.
Then, we normalize each CSI sample using z-score normalization:
\begin{equation}
\tilde{A}(i, k, t) = \frac{A(i, k, t) - \mu}{\sigma + \epsilon},
\end{equation}
where $\mu$ and $\sigma$ denote the mean and standard deviation computed over the entire CSI amplitude samples, and 
$\epsilon$ is a small constant to ensure numerical stability, which is empirically set.

Each CSI amplitude input sample is  represented as a three-dimensional antenna-frequency-time matrix, which is interpreted as a multi-channel image and fed into a CNN.
We adopt a CNN to learn discriminative spatio-temporal patterns from CSI amplitude dynamics. 
The reason why we use a CNN model is because it can jointly capture local temporal transitions and frequency-domain correlations across subcarriers.


Formally, the CNN outputs a probability distribution over predefined four body movement classes:
\begin{equation}
\hat{y} = \arg\max_{c} f_{\theta}\!\left(\{ \tilde{A}(i, k, t) \}\right),
\end{equation}
where $f_{\theta}(\cdot)$ denotes the CNN parameterized by $\theta$, and $c$ indexes the set of four sleep-related movement categories. 


%% file: evaluation.tex
\begin{figure}
    \centering
    \includegraphics[width=0.28\linewidth]{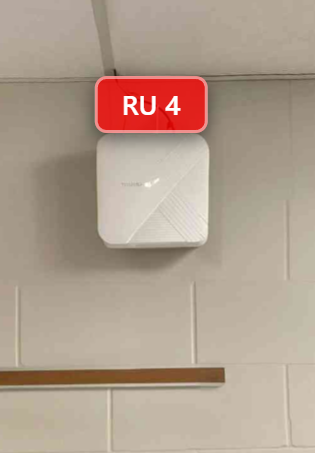}    
    \includegraphics[
    trim = 280 300 140 420,
    clip,
    width=0.7\linewidth,]{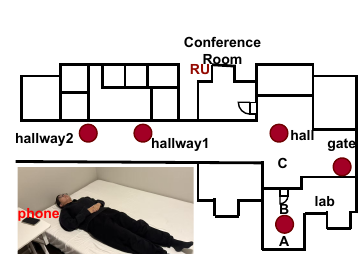}
    \caption{MSU's private 5G network platform.}
    \label{fig:experiment}
\end{figure}
\subsection{Implementation}
Ideally, the proposed solution should be evaluated on an outdoor micro-cell 5G system. However, access to public 5G infrastructure is highly limited. Therefore, we conduct our evaluation on the MSU private 5G network (MSU-P5G \cite{MSU_Private_5G_Network_Platform}). MSU-P5G consists of six indoor commercial 5G base stations and 20 smartphones deployed throughout MSU’s engineering building. Fig.~\ref{fig:experiment} illustrates the 5G radio unit and the corresponding deployment floor plan. Additional details about the private 5G network are available at \url{https://inss.egr.msu.edu/5g}.

The 5G base stations are configured to instruct smartphones to transmit SRS for uplink channel sounding. The base station periodically measures CSI at the distributed unit (DU) and forwards the measurements to an xApp running on the Near-RT RIC via the E2 interface. The xApp then infers respiration and body movements using the collected CSI samples.

\subsection{Experimental Setting and Performance Metrics}

\noindent
\textbf{Experimental Setting.} 
To evaluate our proposed solution under diverse sensing conditions, we conducted experiments at multiple locations as shown in Fig.~\ref{fig:experiment}. 
Specifically, we measured the CSI data at five different locations. 
Subjects performed four representative actions (body turn, sitting up, arm move, and leg move) in a normal way. 

\noindent
\textbf{Evaluation Metrics.} 
For respiration monitoring, estimated breathing rates are compared against reference measurements to evaluate estimation accuracy. 
For sleep movement classification, performance is evaluated using classification accuracy on the test set.


\begin{figure}
    \centering
    \begin{subfigure}[b]{0.475\linewidth}
        \centering
        \includegraphics[width=\linewidth]{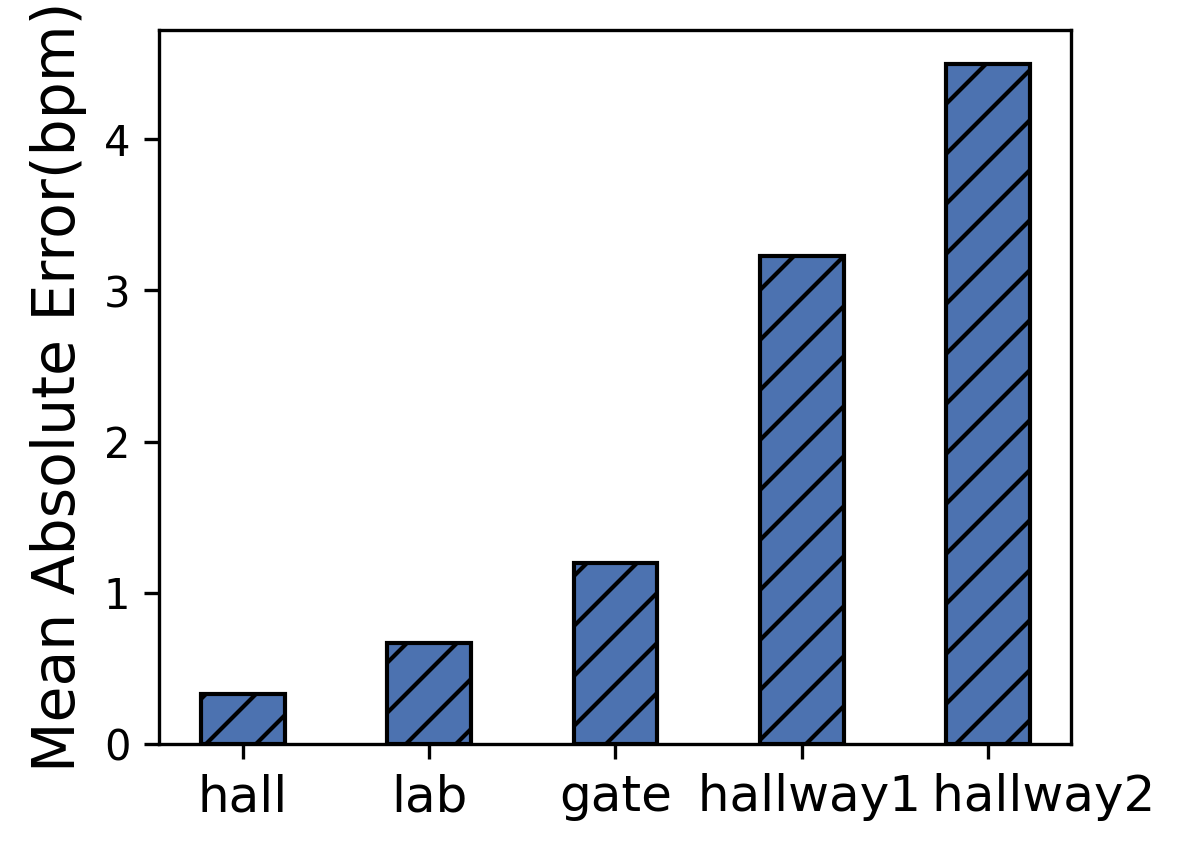}
        \caption{Mean absolute error (bpm).}
        \label{fig:sub1}
    \end{subfigure}
    \hfill
    \begin{subfigure}[b]{0.49\linewidth}
        \centering
        \includegraphics[width=\linewidth]{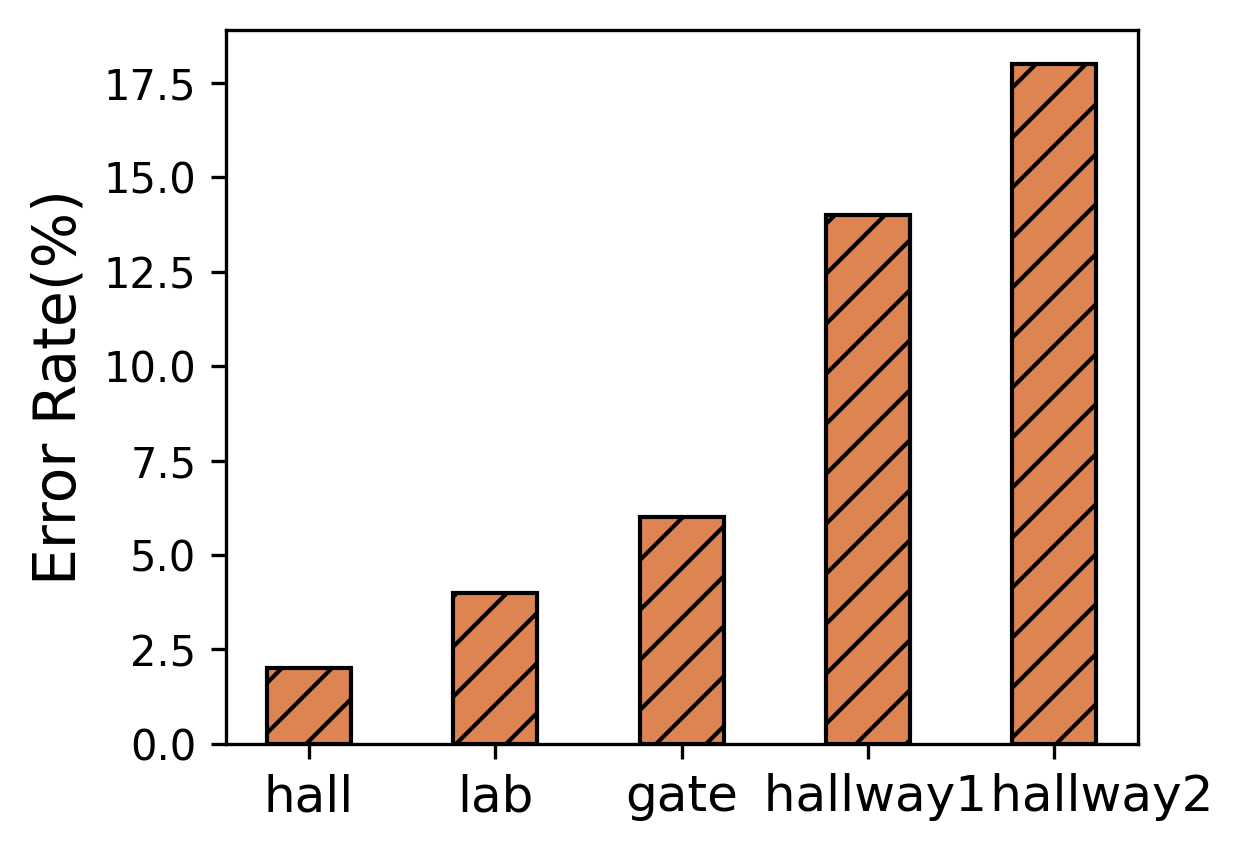}
        \caption{Esti. error percentage (\%).}
        \label{fig:sub2}
    \end{subfigure}

    \caption{Respiration estimation error.}
    \label{fig:respiration}
\end{figure}

\subsection{Respiration Rate Estimation}

\noindent
\textbf{Estimation Accuracy.}
Fig.~\ref{fig:respiration} presents the respiration rate performance across different indoor scenarios and testing locations. 
Over the five locations, the average estimation error is 8.8\%.
The results indicate that our system achieves higher accuracy in the \textit{lab}, \textit{hall}, and \textit{gate} environments, where the mean absolute error remains low and the error rate is consistently stable. 
In contrast, the performance in the \textit{hallway} scenario is comparatively degraded. This degradation can be attributed to the significant path loss of thick concrete wall penetration and multi-wall penetration from the phone to the base station (see Fig.~\ref{fig:experiment}).


\noindent
\textbf{Resilience to Interference.}
Despite environmental variability, the proposed respiration monitoring approach demonstrates a notable degree of robustness against common indoor disturbances. During the experiments, normal human activities such as walking, speaking, and routine work were present in the surrounding environment. When the interference sources were located more than one meter away from the monitored subject, and the mobile device was placed in close proximity to the target, respiration signals could still be reliably extracted. The interference resilience can be attributed to the use of a bandpass filter within the typical respiration frequency range (i.e., 0.1--0.5Hz). 
The interference from most of routine activities is out of this frequency range, making the system robust for respiration estimation.  

\noindent
\textbf{Demonstration.}
A demo video of real-time respiration measurement is available at:
\url{https://inss.egr.msu.edu/5g-sleep}.



\begin{figure}
    \centering
    \begin{subfigure}[b]{0.48\linewidth}
        \centering
        \includegraphics[
        trim = 0 45 40 30,
        clip,
        width=\linewidth]{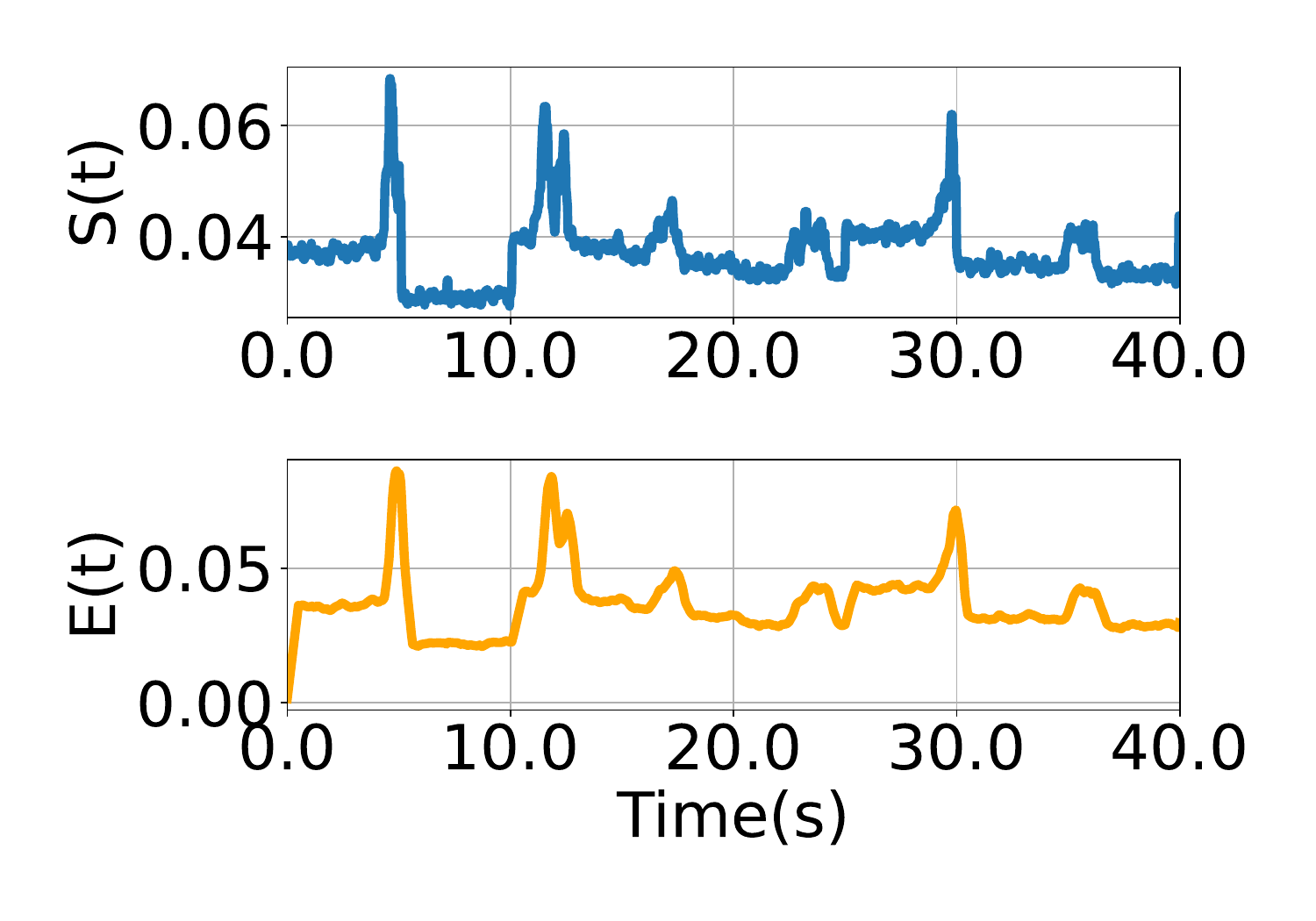}
        \caption{Body turn.}
        \label{fig:turn_over}
    \end{subfigure}
    \hfill
    \begin{subfigure}[b]{0.48\linewidth}
        \centering
        \includegraphics[
        trim = 0 45 40 30,
        clip,
        width=\linewidth]{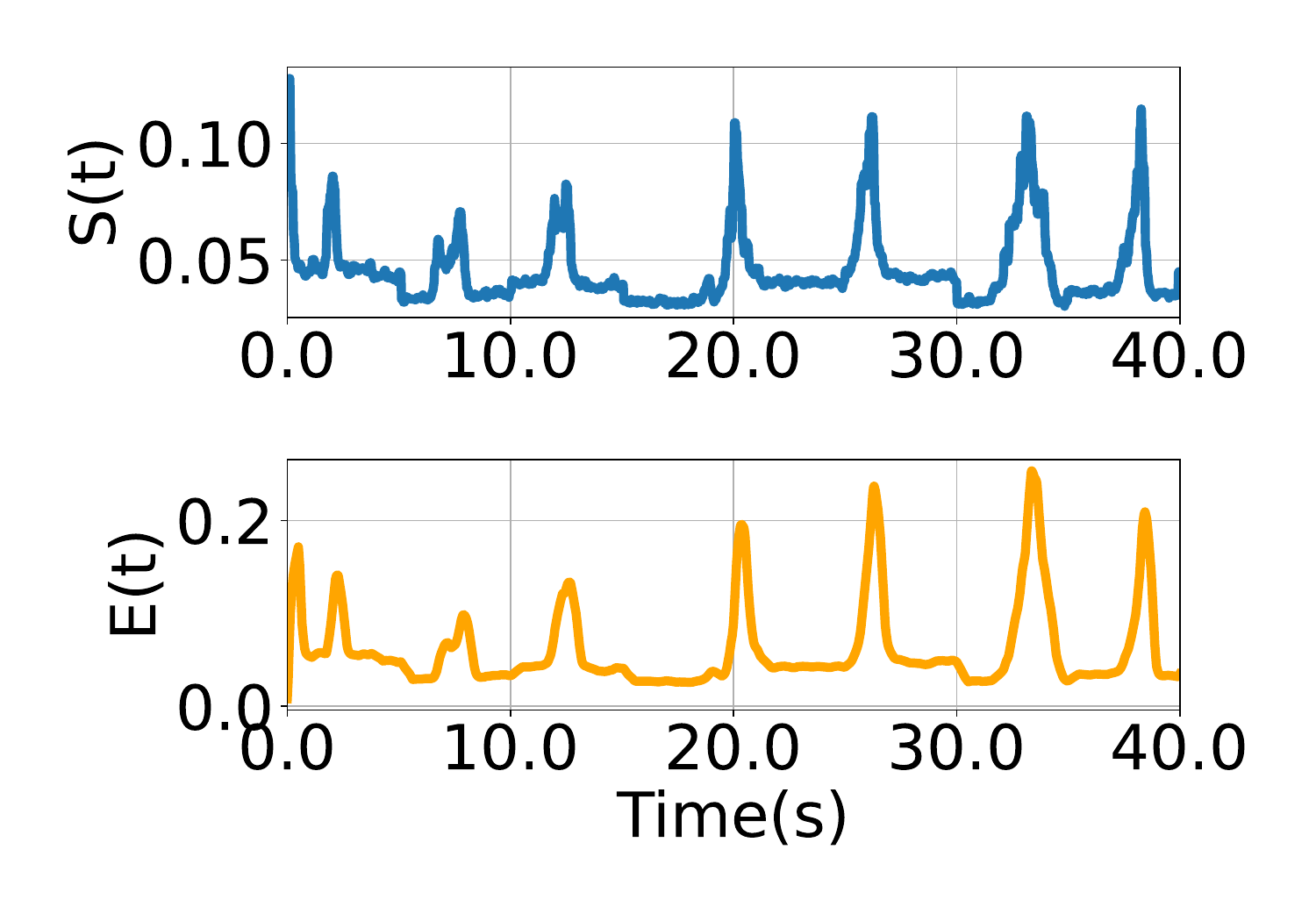}
        \caption{Sitting up.}
        \label{fig:sit_up}
    \end{subfigure}

    \vspace{0.5em}

    \begin{subfigure}[b]{0.48\linewidth}
        \centering
        \includegraphics[
        trim = 0 45 40 30,
        clip,
        width=\linewidth]{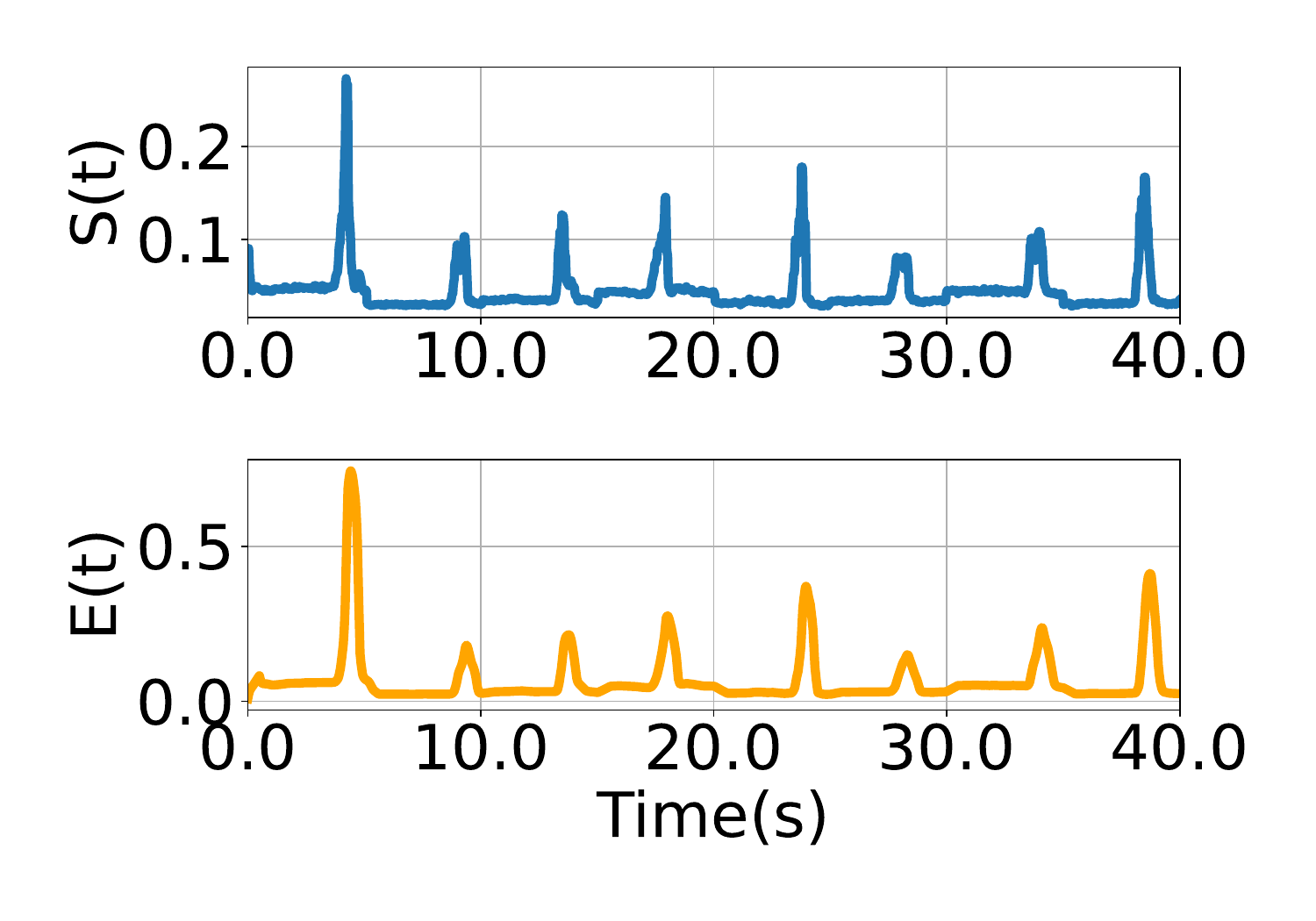}
        \caption{Arm move.}
        \label{fig:arm}
    \end{subfigure}
    \hfill
    \begin{subfigure}[b]{0.48\linewidth}
        \centering
        \includegraphics[
        trim = 0 45 40 30,
        clip,
        width=\linewidth]{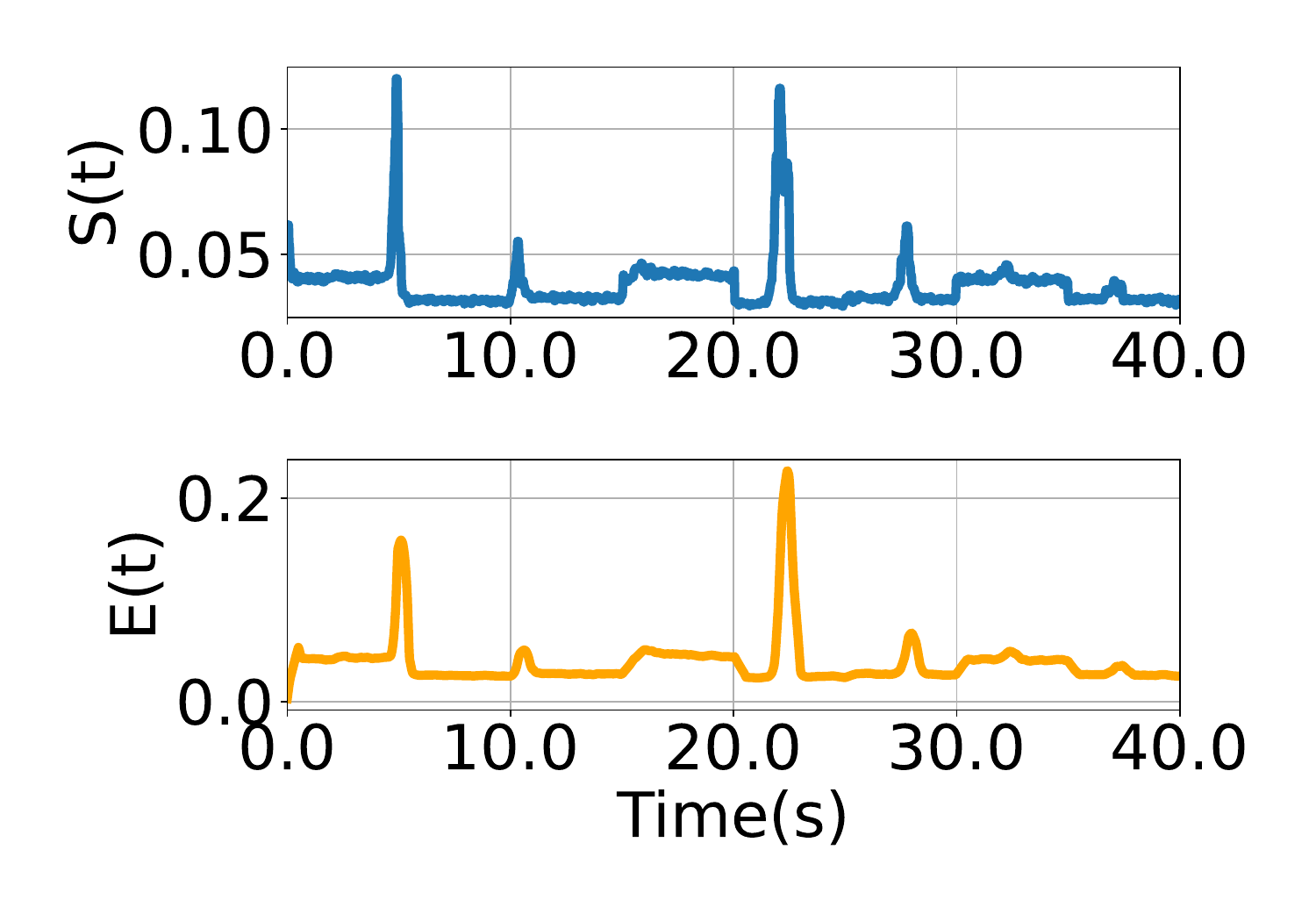}
        \caption{Leg move.}
        \label{fig:leg}
    \end{subfigure}
    \caption{CSI amplitude changes during different movements. }
    \label{fig:movements_2x2}
\end{figure}

\subsection{Body Movement Classification}


\noindent
\textbf{CSI Signature of Movement.}
To understand how the CNN classifier differentiates different sleep movement actions, we first plot the change of CSI amplitude during those movements. 
Specifically, denote $S(t)$ as the mean absolute CSI amplitude change, i.e., 
$S(t) = \sum_{\forall i, k}|A(i, k, t) - A(i, k, t-1)|$.
Denote $E(t) = \frac{1}{W}\sum_{t'=t-W}^t S(t')$ as the short-term energy variation, where $W$ is the window size. 
Fig.~\ref{fig:movements_2x2} plots an instance of $S(t)$ and $E(t)$ during different sleep movements of a subject. 
It can be seen that the movements have clear signature on the CSI amplitude, underpinning the CNN-based classification.

\begin{figure}[!t]
    \centering
    \begin{subfigure}[b]{0.48\linewidth}
        \centering
        \includegraphics[
        width=\linewidth]{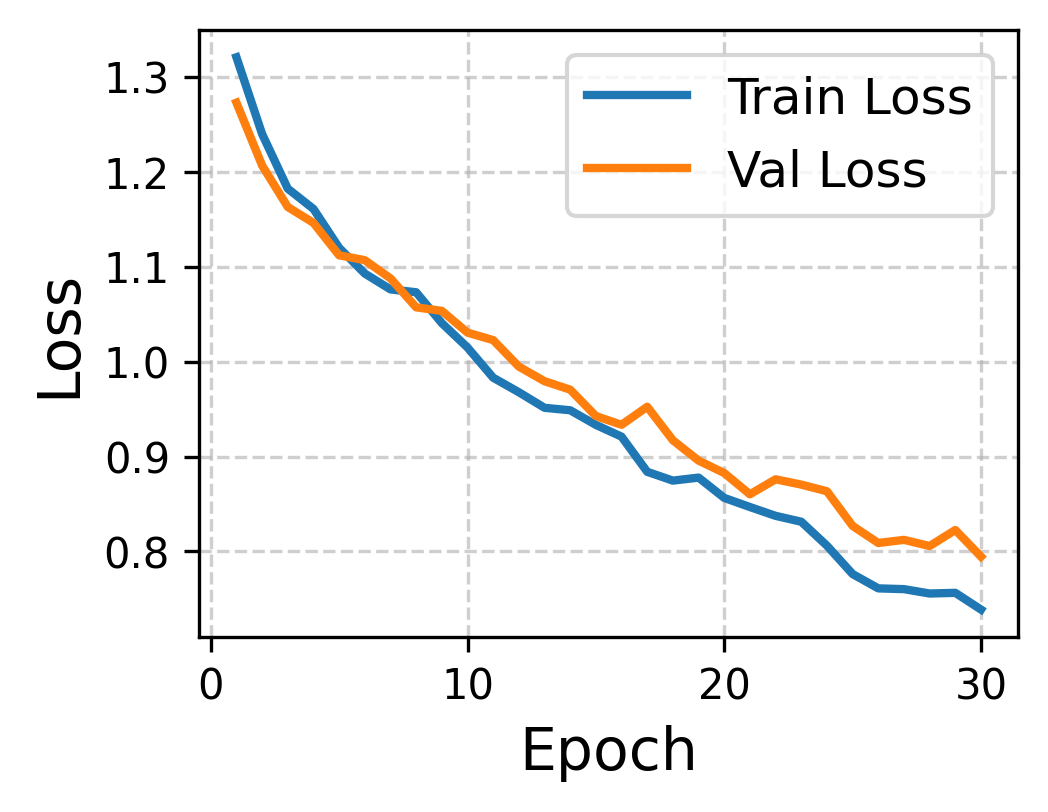}
        \caption{Loss curve.}
        \label{fig:sub1}
    \end{subfigure}
    \hfill
    \begin{subfigure}[b]{0.48\linewidth}
        \centering
        \includegraphics[width=\linewidth]{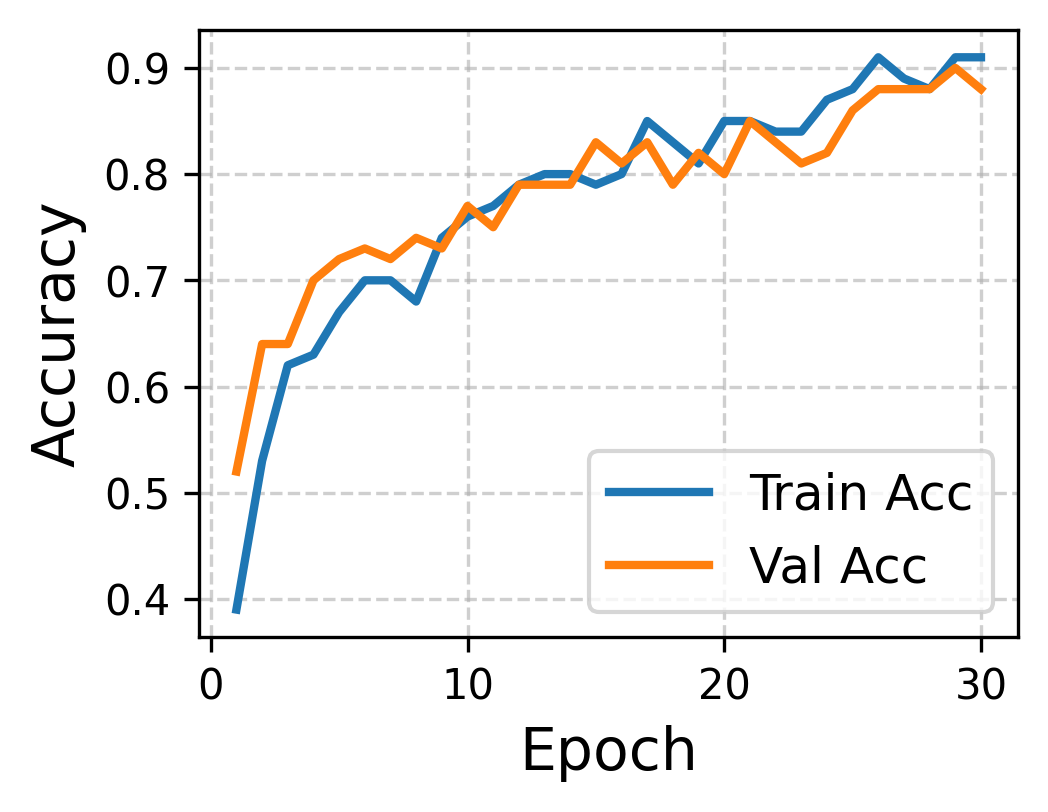 }
        \caption{Accuracy Curve.}
        \label{fig:sub2}
    \end{subfigure}
    \caption{Convergence of CNN-based movement classifier.}
    \label{fig:Training}
\end{figure}

\noindent
\textbf{Training Convergence of CNN.}
To train the CNN classifier, a total of 564 CSI samples were collected from three participants from each of the 5 locations, with each CSI sample corresponding to an annotated label (one of those four categories) as the ground truth. 
Fig.~\ref{fig:Training} presents the convergence performance of the CNN-based sleep movement classifier. 

\begin{figure}[!t]
    \centering
    \begin{subfigure}[b]{0.48\linewidth}
        \centering
        \includegraphics[
        width=\linewidth]{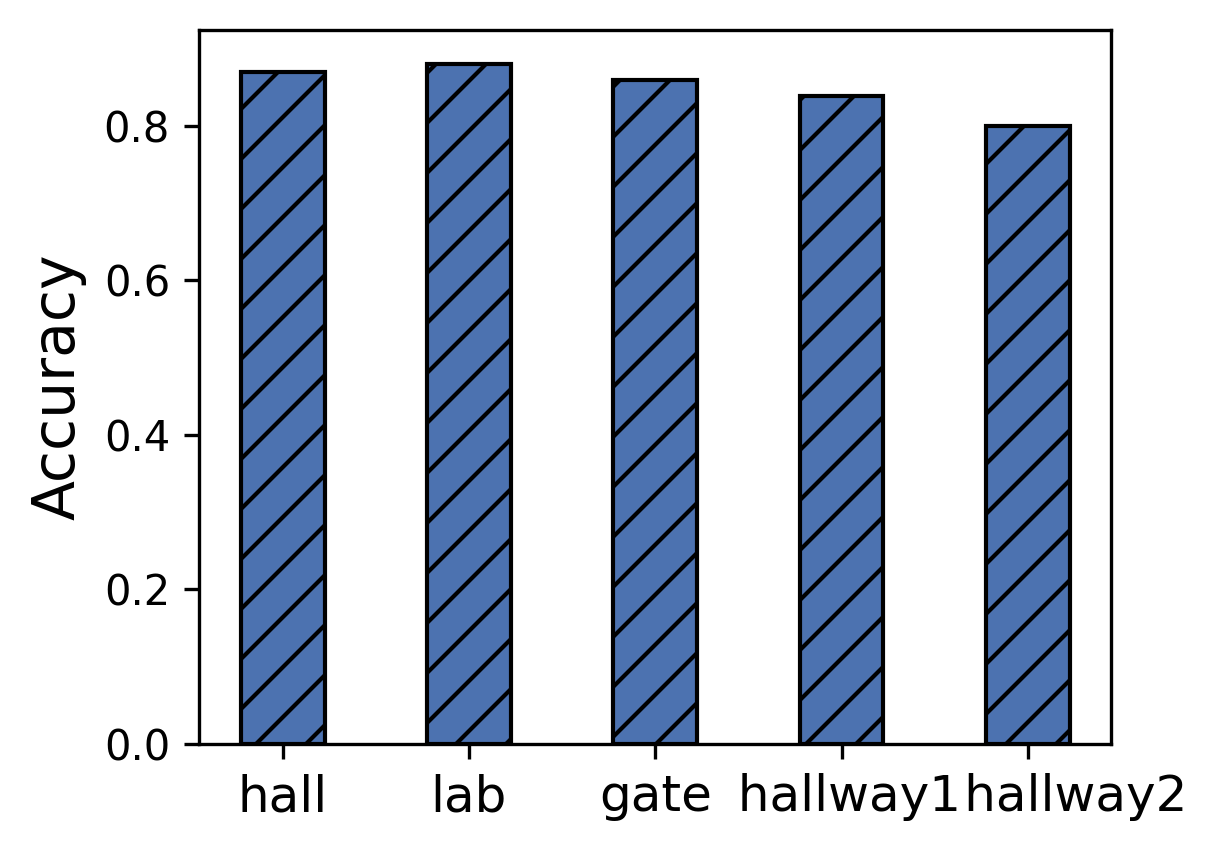}
        \caption{Accuracy Based on Location.}
        \label{fig:sub1}
    \end{subfigure}
    \hfill
    \begin{subfigure}[b]{0.48\linewidth}
        \centering
        \includegraphics[width=\linewidth]{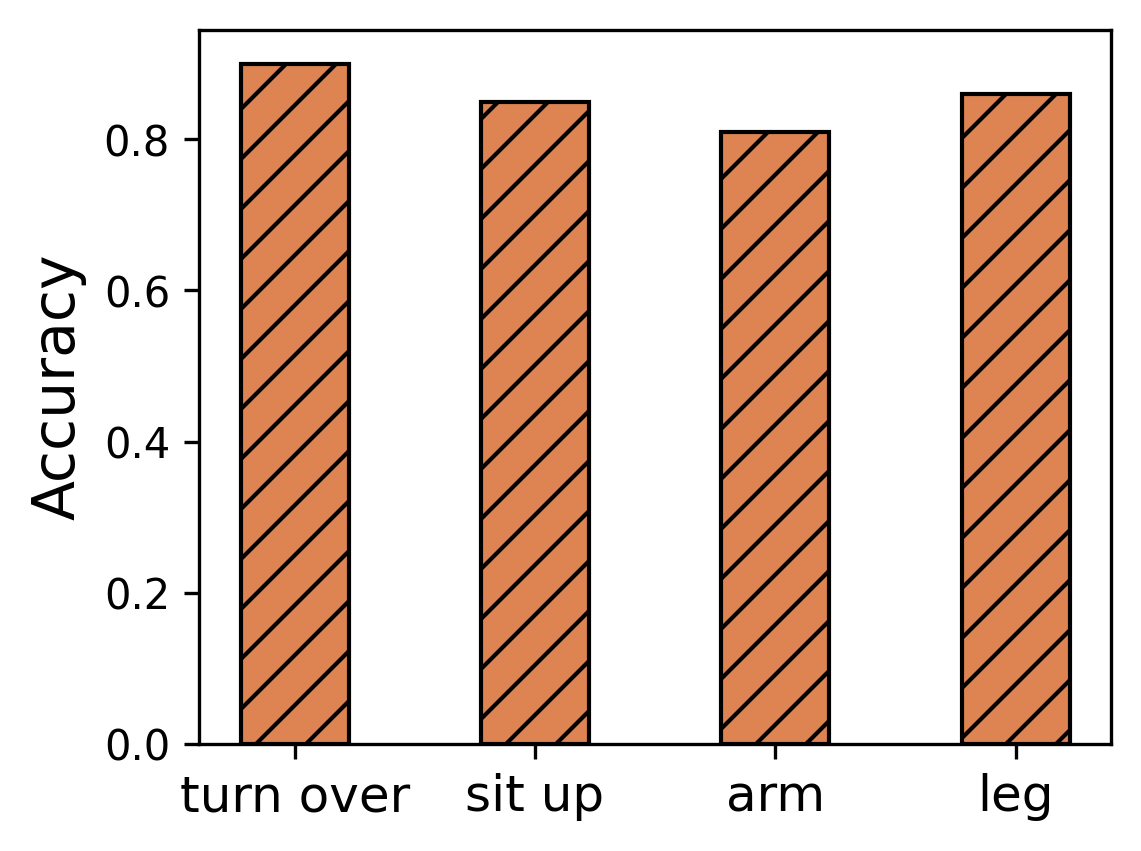 }
        \caption{Accuracy Based on Category.}
        \label{fig:sub2}
    \end{subfigure}
    \caption{Convergence of CNN-based movement classifier.}
    \label{fig:Result}
\end{figure}

\noindent
\textbf{Classification Accuracy.}
We evaluated our model on the held-out test set consisting of 340 samples, evenly distributed across all locations and movements. Fig.~\ref{fig:Result} presents the classification results. 
In Fig.~\ref{fig:Result}(a), the results grouped by location. Classification accuracy remained above 80\% across all locations, demonstrating robust recognition performance even under adverse environmental conditions. 
Fig.~\ref{fig:Result}(b) shows the results grouped by posture. The classification accuracy for all postures remains above 80\%, demonstrating that the model can reliably distinguish the macro movements.  

Overall, these results highlight that while environmental factors and movement amplitude affect the detection rate, our system maintains high classification accuracy across both spatial and activity dimensions.


